\documentclass[aps,prd,12pt,citeautoscript,superscriptaddress,showpacs]{revtex4-1}
\usepackage{xcolor}
\usepackage[colorlinks=true, pdfstartview=FitV,linkcolor=blue, citecolor=blue, urlcolor=blue]{hyperref}

\newcommand{\be}{\begin{equation}}
\newcommand{\ee}{\end{equation}}
\newcommand{\bea}{\begin{eqnarray}}
\newcommand{\eea}{\end{eqnarray}}

\usepackage{psfrag}
\usepackage{epsfig}
\usepackage{latexsym,array,theorem,mathrsfs,subfigure,bm,float}
\usepackage{amsmath}
\usepackage{amsfonts}
\usepackage{amssymb}
\usepackage{bbm}
\usepackage{color}

\renewcommand{\theequation}{\arabic{section}.\arabic{equation}}

\begin{document}

\title{De Sitter vacua in ghost-free massive gravity theory}

\author{Charles Mazuet}
\email{\tt Charles.Mazuet@lmpt.univ-tours.fr}
\affiliation{
Laboratoire de Math\'{e}matiques et Physique Th\'{e}orique CNRS-UMR 7350, \\ 
Universit\'{e} de Tours, Parc de Grandmont, 37200 Tours, France}

\author{Mikhail~S.~Volkov}
\email{\tt volkov@lmpt.univ-tours.fr}
\affiliation{
Laboratoire de Math\'{e}matiques et Physique Th\'{e}orique CNRS-UMR 7350, \\ 
Universit\'{e} de Tours, Parc de Grandmont, 37200 Tours, France}

\affiliation{
Department of General Relativity and Gravitation, Institute of Physics,\\
Kazan Federal University, Kremlevskaya street 18, 420008 Kazan, Russia
}

\begin{abstract} 
\vspace{1 cm}
We present a simple procedure to obtain all de Sitter solutions in the ghost-free
massive gravity theory by using the Gordon ansatz. 
For these solutions the physical metric can be conveniently 
viewed as describing a hyperboloid in 5D Minkowski space, 
while the flat reference metric depends on the Stuckelberg field $T(t,r)$ that satisfies 
the equation  $(\partial_t{T})^2-(\partial_r T)^2=1$. This equation has infinitely 
many solutions, hence there are infinitely many de Sitter vacua with 
different physical properties. Only the simplest solution with $T=t$ 
has been previously studied since it is 
manifestly homogeneous and isotropic, but it is unstable. However, other solutions
could be stable. 
We require the timelike 
isometry to be common for both metrics, and this gives  
physically distinguished solutions since only for them the canonical energy 
is time-independent. We conjecture that these solutions minimize the energy  
and are therefore stable. 
We also show that in some cases solutions can be homogeneous and isotropic
in a non-manifest way such that their symmetries are not obvious. 
All of this suggests that the theory may admit viable cosmologies.

\end{abstract} 

\maketitle

\section{Introduction}
\setcounter{equation}{0} 

The discovery of the ghost-free massive gravity theory 
by de Rham, Gabadadze, and Tolley (dRGT) \cite{deRham:2010kj} 
(see \cite{Hinterbichler:2011tt,*deRham:2014zqa} for a review)
opens up the possibility to explain the dark energy and the 
cosmic acceleration 
\cite{1538-3881-116-3-1009,*0004-637X-517-2-565} by a tiny 
mass of the gravitons. The dRGT field equations 
admit the de Sitter solution with the cosmological constant mimicked by the 
graviton mass. This solution can describe the late time acceleration, but 
a special analysis is needed to decide 
whether its other  properties  are physically acceptable.  

A closer look reveals that the de Sitter solution in the dRGT theory is actually not unique,
and a number of its versions have been found
\cite{Koyama:2011xz,*Koyama:2011yg},%
\cite{Chamseddine:2011bu},%
\cite{D'Amico:2011jj},%
\cite{Gumrukcuoglu:2011ew},%
\cite{Gratia:2012wt,*Wyman:2012iw},
\cite{Volkov:2012cf,*Volkov:2012zb,*Volkov:2013roa}. 
A special attention was received by one solution 
whose physical and reference metrics are of the manifestly homogeneous and isotropic 
Freedman-Lema$\hat{{\i}}$tre-Robertson-Walker 
(FLRW) form \cite{Gumrukcuoglu:2011ew}. However,
a detailed analysis revealed that this solution is unstable 
\cite{DeFelice:2012mx,*DeFelice:2013awa}. 
For other known solutions only the physical metric is manifestly FLRW 
while the reference metric looks inhomogeneous, for which reason  
they are considered to be less interesting  \cite{D'Amico:2011jj}.
All of this has reduced the interest towards the dRGT theory,
the focus shifting towards its extensions, as for example the bigravity 
\cite{Hassan:2011zd},\cite{Volkov:2011an,*vonStrauss:2011mq,*Comelli:2011zm},%
\cite{Akrami:2012vf,*Comelli:2014bqa,*Konnig:2014xva,%
*DeFelice:2014nja,*Solomon:2014iwa} and other generalizations 
admitting FLRW solutions 
\cite{D'Amico:2012zv,*Huang,*DeFelice:2013dua,*Mukohyama:2014rca,*deRham:2014gla,*Heisenberg:2015voa}.

However, we would like to argue that it may be premature to abandon the dRGT theory
on the basis of negative evidence obtained from just one solution, because 
the theory admits infinitely many other solutions that could 
be physically interesting. They all have the same physical  (de Sitter) metric
but different values of the reference metric depending on the
Stuckelberg field $T(t,r)$ subject to a complicated differential equation 
\cite{D'Amico:2011jj},\cite{Gratia:2012wt},%
\cite{Volkov:2012cf,*Volkov:2012zb,*Volkov:2013roa}.
Below we shall describe a simple way to obtain these solutions by applying the  
Gordon ansatz \cite{Visser} and using the global embedding coordinates. 
The $T$-equation then assumes a simple form, 
$(\partial_t{T})^2-(\partial_r T)^2=1$, whose essentially general solution is known.  
The simplest solution $T=t$ 
\cite{Gumrukcuoglu:2011ew} is unstable 
\cite{DeFelice:2012mx,*DeFelice:2013awa}
but other solutions could be stable. 
One can choose $T$ such that both metrics are invariant under the timelike  
isometry, which gives special solutions since only for them 
the canonical energy is time independent.
We conjecture that their energy is minimal and hence  these solutions are stable. 
We also give explicit examples where the reference metric 
looks inhomogeneous but shares with the physical metric the same  
translational and rotational isometries. Hence, solutions considered to be non-FLRW 
can actually be homogeneous and isotropic. 
All of this suggests that viable  dRGT cosmologies  
may exist.

\section{The dRGT massive gravity}
\setcounter{equation}{0} 

The theory is  defined on a four-dimensional spacetime manifold endowed with 
two Lorentzian metrics, the physical metric $g_{\mu\nu}$ 
and the flat reference metric  
$
f_{\mu\nu}=\eta_{AB}\partial_\mu\Phi^A\partial_\nu\Phi^B
$
with $\eta_{AB}={\rm diag}[-1,1,1,1]$. 
The fields $\Phi^A(x)$ are sometimes called 
Stuckelberg scalars. The theory is defined 
by the action 
\bea                                      \label{1}
&&S
=\frac{M_{\rm Pl}^2}{m^2}\int \, \left(
\frac{1}{2}\,R({g})
-{\cal U}\right)\sqrt{-g}\,d^4x
\,.
\eea
 The metrics and all coordinates are 
assumed to be dimensionless, the length scale
being the inverse graviton mass $1/m$.  
 The interaction between the two metrics 
is expressed by  
 a scalar function 
of the tensor 
$
\gamma^\mu_{~\nu}=\sqrt{{{g}}^{\mu\alpha}{{f}}_{\alpha\nu}}
$
where ${{g}}^{\mu\nu}$ is the inverse of ${g}_{\mu\nu}$
and the square root
is understood in the matrix sense, i.e. 
$
(\gamma^2)^\mu_{~\nu}\equiv \gamma^\mu_{~\alpha}\gamma^\alpha_{~\nu}={g}^{\mu\alpha}
{f}_{\alpha\nu}.
$
If $\lambda_A$ ($A=0,1,2,3$) are the eigenvalues of $\gamma^\mu_{~\nu}$
then the interaction potential is given by 
$
{\cal U}=\sum_{n=0}^4 b_k\,{\cal U}_k
$
where $b_k$ are parameters and 
${\cal U}_k$ are defined 
by the 
relations
\bea                        \label{4}
{\cal U}_0&=&1,~~~~~
{\cal U}_1=
\sum_{A}\lambda_A=[\gamma],~~~~~
{\cal U}_2=
\sum_{A<B}\lambda_A\lambda_B 
=\frac{1}{2!}([\gamma]^2-[\gamma^2]),\nonumber \\
{\cal U}_3&=&
\sum_{A<B<C}\lambda_A\lambda_B\lambda_C
=
\frac{1}{3!}([\gamma]^3-3[\gamma][\gamma^2]+2[\gamma^3]),\nonumber \\
{\cal U}_4&=&
\lambda_0\lambda_1\lambda_2\lambda_3
=
\frac{1}{4!}([\gamma]^4-6[\gamma]^2[\gamma^2]+8[\gamma][\gamma^3]+3[\gamma^2]^2
-6[\gamma^4])\,. \nonumber 
\eea
Here, using the hat to denote matrices, one defines 
$[\gamma]\equiv {\rm tr}(\hat{\gamma})= \gamma^\mu_{~\mu}$, 
$[\gamma^k]\equiv {\rm tr}(\hat{\gamma}^k)= (\gamma^k)^\mu_{~\mu}$. 
The parameters $b_k$ can apriori be arbitrary, but 
if one requires the flat space to be a solution of the theory and 
$m$ to be the Fierz-Pauli
mass of the gravitons in flat space,  then 
the five $b_k$ are 
expressed in terms of two arbitrary 
parameters, sometimes called $c_3$ and $c_4$, as 
\be                \label{bbb}
b_0=4c_3+c_4-6,~~
b_1=3-3c_3-c_4,~~
b_2=2c_3+c_4-1,~~
b_3=-(c_3+c_4),~~
b_4=c_4.
\ee
The metric $g_{\mu\nu}$ and the scalars $\Phi^A$ are the variables of the theory.
Varying the action (\ref{1}) with respect to $g_{\mu\nu}$
gives the Einstein equations 
$G_{\mu\nu}=T_{\mu\nu}$
with 
\bea                                \label{T0}
T^\mu_{~\nu}&=&
\{b_1\,{\cal U}_0+b_2\,{\cal U}_1+b_3\,{\cal U}_2
+b_4\,{\cal U}_3\}\gamma^\mu_{~\nu} 
-\{b_2\,{\cal U}_0+b_3\,{\cal U}_1+b_4\,{\cal U}_2\}(\gamma^2)^\mu_{~\nu} \nonumber  \\
&+&\{b_3\,{\cal U}_0+b_4\,{\cal U}_1\}(\gamma^3)^\mu_{~\nu} 
-b_4\,{\cal U}_0\,(\gamma^4)^\mu_{~\nu}-{\cal U}\,\delta^\mu_\nu \,.
\eea
Varying the action with respect to $\Phi^A$ gives 
the conservation conditions  
$
{\nabla}_\mu T^\mu_\nu=0\,.
$
These are equations for the Stuckelberg scalars, but they 
are actually not independent and  
follow from the Bianchi identities for the Einstein equations.

\section{de Sitter space}
\setcounter{equation}{0} 
The above field equations admit solutions for which the physical metric is de Sitter. 
Specifically, 
the de Sitter space can be globally visualized 
as the hyperboloid 
\be
-X_0^2+\sum_i X_i^2+X_4^2=\alpha^2
\ee
is the 5D Minkowski space with the metric
\be                                  \label{dSS}
ds^2=-dX_0^2+\sum_i dX_i^2+dX_4^2.
\ee 
Rescaling the coordinates, 
$X_0=\alpha t$, $X_i=\alpha x_i$, $X_4=\alpha r$
with $x_i\equiv (x,y,z)$, 
the metric reads 
\bea                                  \label{dS}
ds_g^2&=&\alpha^2\left\{-dt^2+dr^2+dx^2+dy^2+dz^2
\right\}  
\nonumber \\
&=&\alpha^2\left\{-dt^2+dr^2+dR^2+R^2\,d\Omega^2
\right\} 
\eea 
where $d\Omega^2=d\vartheta^2+\sin^2\vartheta d\varphi^2$ and 
\be                                \label{hyper}
R^2\equiv x^2+y^2+z^2=1+t^2-r^2.
\ee
Let us choose the flat 
reference metric as
\be                        \label{dSf}
ds_f^2=\alpha^2 u^2\left\{-dT^2+dX^2
+dY^2+dZ^2 \right\}, 
\ee 
where $u$ is a constant and $T,X,Y,Z$ are the Stuckelberg  fields. 

It turns out that $\alpha$, $u$ and $T,X,Y,Z$ 
can be chosen such that the two metrics  fulfill the field equations. 
Specifically, it is sufficient  to make sure that 
they fulfill the following relation  (the Gordon ansatz) \cite{Visser},
\be                            \label{Gordon} 
f_{\mu\nu}=\omega^2\left(g_{\mu\nu}+(1-\zeta^2)V_\mu V_\nu\right),
\ee
where $\omega,\zeta$ are functions and 
\be                              \label{norm}
g^{\mu\nu}V_\mu V_\nu\equiv V^\mu V_\mu=-1.
\ee
If Eq.\eqref{Gordon} is fulfilled, 
then one can see at once that 
 \be 
\gamma^\mu_{~\nu}
=
\sqrt{g^{\mu\alpha}f_{\alpha\nu} }
=\omega\left(
\delta^\mu_\nu+(1-\zeta)V^\mu V_\nu
\right),
\ee
since $\gamma^\mu_\alpha\gamma^\alpha_\nu=g^{\mu\alpha}f_{\alpha\nu}$. 
One has 
$
(\gamma^n)^\mu_{~\nu}=\omega^n\left(
\delta^\mu_\nu+(1-\zeta^n)V^\mu V_\nu
\right)
$
and so 
the energy-momentum tensor \eqref{T0} is 
\be                           \label{TT}
T^\mu_\nu=-\left\{P_0(\omega)-\zeta\omega P_1(\omega)\right\}\delta^\mu_\nu
+\omega(\zeta-1)P_1(\omega)V^\mu V_\nu\,
\ee
with 
\be
P_m(\omega)= b_m+2b_{m+1}\,\omega+b_{m+2}\,\omega^2;~~~~~m=0,1,2. 
\ee
Let us set $\omega=u$ where $u$ is a constant chosen such that 
\be                         \label{u}
P_1(u)=0.
\ee 
Then
the energy-momentum tensor \eqref{TT} reduces to   
$T^\mu_\nu=-P_0(u)\delta^\mu_\nu$
and the 
Einstein equations 
become 
\be
G^\mu_\nu+\Lambda\delta^\mu_\nu=0
\ee
with $\Lambda=P_0(u)$.
The de Sitter metric \eqref{dS} is a solution of these equations provided that 
\be                         \label{alpha}
\frac{1}{\alpha^2}=\frac{\Lambda}{3}=\frac{P_0(u)}{3}. 
\ee
Therefore, the metrics \eqref{dS} and \eqref{dSf} will indeed fulfill the 
field equations if $u$ and $\alpha$ are defined by \eqref{u},\eqref{alpha},
provided that one can adjust 
the functions $T,X,Y,Z$ such that the Gordon relation \eqref{Gordon}
is fulfilled. 

Let us choose in \eqref{dSf} $T=T(t,r)$, $X=x$, $Y=y$, $Z=z$ so that the f-metric becomes   
\be                        \label{dSff}
ds_f^2=\alpha^2 u^2\left\{-dT^2+dx^2
+dy^2+dz^2 \right\}
=
\alpha^2 u^2\left\{-dT^2+dR^2
+R^2 d\Omega^2 \right\}.
\ee 
The two metrics 
\eqref{dS} and \eqref{dSff} are related to each other as
\be 
ds_f^2=u^2\left(
ds_g^2+dt^2-dr^2-dT^2
\right).
\ee
This will be compatible with the Gordon relation \eqref{Gordon} if
\be                            \label{grd}
\partial_\mu t\partial_\nu t-
\partial_\mu r\partial_\nu r-
\partial_\mu T\partial_\nu T=(1-\zeta^2)V_\mu V_\nu\,. 
\ee 
Assuming that the indices $\mu,\nu$ correspond to $(t,r,\vartheta,\varphi)$ 
yields $V_\vartheta=V_\varphi=0$ and 
\bea                       \label{TTT}
(\partial_t T)^2-1=(\zeta^2-1)V_t^2\,,  \nonumber \\
(\partial_r T)^2+1=(\zeta^2-1)V_r^2\,,  \nonumber \\
\partial_t{T}\,\partial_r T=(\zeta^2-1)V_t V_r\,. 
\eea
From the first two of these relations one obtains 
\be                          \label{VV}
V_t^2=\frac{ (\partial_t {T})^2-1}{\zeta^2-1},~~~~~~
V_r^2=\frac{ (\partial_ r T)^2+1}{\zeta^2-1},
\ee
while the normalization condition \eqref{norm} determines $\zeta$. 
Finally, inserting \eqref{VV} to the third relation in \eqref{TTT} yields 
\bea 
(\partial_t {T})^2 (\partial_r T)^2 = 
( (\partial_t {T})^2-1)( (\partial_r T)^2+1  )  
\eea
and therefore 
\be                          \label{T}
(\partial_t{T})^2-(\partial_r T)^2=1. 
\ee 
This completes the procedure, because $V_\mu$ and $\zeta$ are determined by the above 
formulas and 
the Gordon relation is fulfilled.  

Summarizing, the de Sitter solution in the theory is described by 
Eqs.\eqref{dS},\eqref{dSff} where $u$,$\alpha$ are defined by \eqref{u},\eqref{alpha}
and $T$ is a solution of the differential equation \eqref{T}. 
Since there are infinitely many $T$'s subject to \eqref{T}, 
there are infinitely many de Sitter solutions. They all
have the same physical metric  \eqref{dS} but differ one from the other 
by the choice of $T$ in the reference metric \eqref{dSff}. 
The physical properties of solutions with different $T$'s,
as for example their stability, can be different. 

These solutions were actually obtained previously 
\cite{D'Amico:2011jj},\cite{Gratia:2012wt},\cite{Volkov:2012cf,*Volkov:2012zb,*Volkov:2013roa}, 
but within a different computation scheme yielding  the $T$-equation in a 
form that gives little hope to solve it (see Eq.\eqref{c0} in the Appendix).  
Our procedure yields its equivalent form  
\eqref{T}, which is simple 
and admits a general solution.   
In addition, by slightly modifying our procedure, we can obtain new things. 
Specifically, 
it was assumed in the above derivation 
 that both metrics have the same  spatial $SO(3)$ symmetry. 
However, let us rather choose 
\be                        \label{dSfff1}
ds_f^2=\alpha^2 u^2\left\{-dt^2+dx^2
+dy^2+dZ^2 \right\} 
\ee 
with $Z=Z(r,z)$, so that the two metrics share the same $SO(1,2)$ symmetry in the 
$t,x,y$ subspace. 
Repeating the above analysis one obtains 
\be                          \label{Ta1}
(\partial_r{Z})^2+(\partial_z Z)^2=1,
\ee  
and this  gives new solutions. 
When expressed in the standard spherical coordinates, their f-metric will not look spherically 
symmetric, since for generic $Z$ it has no common with the g-metric $SO(3)$ symmetry,
although it has its own $SO(3)$ in the $x,y,Z$ space. 
Below
we shall mainly be discussing equation \eqref{T} since the analysis of \eqref{Ta1}
is similar.

\section{The simplest solution}
\setcounter{equation}{0}

Even though there are infinitely many solutions of Eq.\eqref{T}, 
almost all known dRGT cosmologies reported
in the literature correspond to just one simplest solution, 
\be                             \label{sol}
T=t. 
\ee
A slightly more general choice is 
\be                                \label{sol0}
T=\cosh(\xi)\,t+\sinh(\xi)\, r
\ee
with a constant $\xi$. However, the value of $\xi$ can be changed 
by a boost in the $t,r$ plane of the ambient  5D Minkowski space, 
which does not affect the g-metric \eqref{dS}, hence one can  set
$\xi=0$ without loss of generality.
Rewriting \eqref{sol} in different coordinates gives results which
look very different, and it has not been recognized that they actually
describe 
the same solution. Let us see what happens when this solution 
is expressed in the standard 
spatially flat, closed, and open coordinate systems.

\subsubsection{Flat slicing}
Let us express $t,r$,$R$ in \eqref{dS} in terms of 
two new coordinates $\tau$ and $\rho$ as  
\be
t=\sinh\tau+\frac{\rho^2}{2}e^\tau,~~~
r=\cosh\tau-\frac{\rho^2}{2}e^\tau,~~~
R=e^\tau\rho\,.
\ee
This solves the constraint \eqref{hyper} and transforms the de Sitter metric \eqref{dS} 
to the standard FLRW form with flat spatial sections,  
\be                                      \label{flat} 
ds_g^2=\alpha^2\{-d\tau^2+a^2(\tau)(d\rho^2+\rho^2d\Omega^2)\}, 
\ee
where $a(\tau )=e^\tau$. The function $T=t$ can be represented as
\be
T=\frac12\int\frac{d\tau}{\dot{a}(\tau)}+\frac12\left(1+\rho^2 \right)a(\tau). 
\ee
This solution was found in Ref.\cite{D'Amico:2011jj}
for $b_k$ given by \eqref{bbb} with $c_3=c_4=0$,  and later   
for arbitrary $b_k$ \cite{Volkov:2012cf,*Volkov:2012zb,*Volkov:2013roa}
(solution in \cite{D'Amico:2011jj},\cite{Volkov:2012cf,*Volkov:2012zb,*Volkov:2013roa}
contains an integration constant that can be obtained by using \eqref{sol0} instead of \eqref{sol}). 
Although the g-metric \eqref{flat} 
is manifestly homogeneous and isotropic, the f-metric \eqref{dSff}, when expressed in the 
$\tau,\rho$ 
coordinates, becomes non-diagonal and $\rho$-dependent, which suggests that it is inhomogeneous.  
For this reason it is sometimes said   that the dRGT theory does not 
admit genuinely homogeneous and isotropic cosmologies with flat spatial sections 
\cite{D'Amico:2011jj}. 
However, we shall shortly comment on this. 

\subsubsection{Closed slicing}
If one chooses in \eqref{dS} 
\be
t=\sinh(\tau),~~~r=\cosh(\tau)\cos(\rho),~~~~R=\cosh(\tau)\sin(\rho),
\ee
this solves the constraint \eqref{hyper} and the de Sitter metric \eqref{dS}
assumes the FLRW form with closed spatial sections, 
\be
ds_g^2=\alpha^2\{-d\tau^2+a^2(\tau)(d\rho^2+\sin^2(\rho)d\Omega^2)\}
\ee
with $a(\tau)=\cosh(\tau)$. These coordinates
cover the whole of de Sitter space. 
The Stuckelberg field 
is
$
T=\sinh(\tau),
$
and the f-metric \eqref{dSff} expressed in the $\tau,\rho$ coordinates 
is again non-diagonal and $\rho$-dependent, which suggests that there are no 
genuinely homogeneous and isotropic cosmologies with closed spatial sections either. 

\subsubsection{Open slicing}

For the open slicing one has 
\be
t=\sinh(\tau)\cosh(\rho),~~~~r=\cosh(\tau),~~~R=\sinh(\tau)\sinh(\rho), 
\ee
and the g-metric becomes
\be
ds_g^2=\alpha^2\{-d\tau^2+a^2(\tau)(d\rho^2+\sinh^2(\rho)d\Omega^2)\}
\ee
with $a(\tau)=\sinh(\tau)$. The Stuckelberg field is 
$
T=\sinh(\tau)\cosh(\rho)
$ 
and the specialty now is that 
the f-metric \eqref{dSff} becomes diagonal in the $\tau,\rho$ coordinates, 
\be                                      \label{muk}
ds_f^2=\alpha^2 u^2\{-\cosh(\tau)^2d\tau^2+a^2(\tau)(d\rho^2+\sinh^2(\rho)d\Omega^2)\}.
\ee
This solution, discovered in Ref.\cite{Gumrukcuoglu:2011ew}, 
is broadly viewed  as the only homogeneous and 
isotropic dRGT cosmology, because both metrics 
are manifestly homogeneous and isotropic, so that they share the same 
rotational and translational Killing symmetries. 
However, this solution is completely equivalent to its  
flat and closed versions. Therefore, the latter also have the same 
common isometries, hence they are all
homogeneous and isotropic, although their symmetries  are not manifest. 
The conclusion is that sometimes solutions can be FLRW in a non-manifest way. 

At the same time, although homogeneous and isotropic,
the solution $T=t$ is not static whereas the de Sitter space is.
Specifically, 
let us consider the

\subsubsection{Static slicing}
Setting
\be                                          \label{ss}
t=\sqrt{1-\rho^2}\,\sinh(\tau),~~~r=\sqrt{1-\rho^2}\,\cosh(\tau),~~~R=\rho
\ee
solves the condition \eqref{hyper} and reduces the de Sitter metric \eqref{dS} to
the static form 
\be
ds_g^2=\alpha^2\left\{
-\Sigma\,d\tau^2+\frac{d\rho^2}{\Sigma}+\rho^2d\Omega^2
\right\}
\ee
with $\Sigma=1-\rho^2$. 
The $T=t$ solution then becomes 
\be                                     \label{sol1}
T(\tau,\rho)=\sqrt{1-\rho^2}\,\sinh(\tau), 
\ee
and it is non-static even in static coordinates. Therefore, 
the g-metric is invariant 
under the action of the (locally) timelike Killing vector $\partial/\partial\tau$, 
but the Stuckelberg field $T$ and the f-metric are not invariant. 
As a result, the timelike isometry is not shared by both
metrics.

As the solution $T=t$ is not static, it is unlikely to 
describe the ``ground state'' of the theory. This is probably the reason 
why this solution was found to be unstable \cite{DeFelice:2012mx,*DeFelice:2013awa}. 
Therefore, we need to consider  other solutions for $T$.

\section{Other solutions}
\setcounter{equation}{0} 

Solutions of the $T$-equation 
$(\partial_t{T})^2-(\partial_r T)^2=1$ can be constructed in different ways.
A fairly general 
solution 
containing an arbitrary function $W(\xi)$ is given by \cite{Courant},
\bea
T&=&\cosh(\xi)\,t+\sinh(\xi)\,r+W(\xi)\,, \nonumber \\
0&=&\sinh(\xi)\,t+\cosh(\xi)\,r+\frac{dW(\xi)}{d\xi}\,,
\eea
where the second line implicitly determines the dependence of 
$\xi$ on $t,r$.  Together with \eqref{sol0}, this gives if not all but 
probably almost all solutions. However, this formula is difficult to use 
since one cannot explicitly determine  $\xi(t,r)$ for a generic $W(\xi)$.  

\begin{figure}[h]
\hbox to \linewidth{ \hss

	\resizebox{6cm}{5cm}
	{\includegraphics{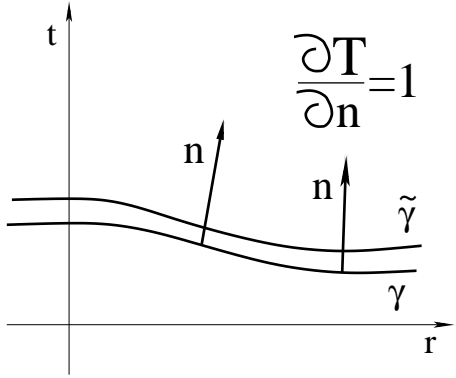}}
	
\hspace{1mm}
\hss}
\caption{{\protect\small 
The $T$-equation in the method of characteristics. 
 }}%
 \label{Fig}
\end{figure} 
The $T$-equation can also be integrated by applying the method of 
characteristics \cite{Courant}, which has  a simple geometric 
interpretation.  Let us consider the 2D Minkowski space spanned by 
$x^a\equiv \{t,r\}$ 
with the metric $g_{ab}={\rm diag}[1,-1]$. The $T$-equation reads 
$g^{ab}\partial_a T\partial_b T\equiv \langle \nabla T,\nabla T\rangle =1$.  
Let $\gamma=x^a(s)$ be a spacelike curve 
and $T$ is constant along it. At every  point of $\gamma$
there is a unit timelike normal $n$ such that $\langle n,n\rangle=1$ and 
$\langle n,\partial/\partial s\rangle =0$. 
The $T$-equation is equivalent to 
$\partial T/\partial n=1$ \cite{Courant}.

This allows one to pass from $\gamma$ where $T=T(\gamma)$ to a neighboring 
curve $\tilde{\gamma}$ where $T=T(\tilde{\gamma})$ (see Fig.\ref{Fig}) and so on,
thereby extending $T$ 
to the whole of the space. The solution is therefore defined, up to an additive constant, 
by the choice of the initial curve $\gamma$. For example, the solution \eqref{sol0}
can be obtained by choosing $\gamma$ to be a straight line.

In practice solutions of $(\partial_t{T})^2-(\partial_r T)^2=1$  can be obtained
by changing the variables and then separating them \cite{Khosravi:2013axa}. 
Let us illustrate the method 
by passing to the static 
coordinates \eqref{ss}, in which case the $T$-equation becomes 
\be                                       \label{sol11}
\frac{1}{\Sigma}\left(\frac{\partial T}{\partial \tau}\right)^2
-
\frac{\Sigma}{1-\Sigma}\left(\frac{\partial T}{\partial \rho}\right)^2=1.
\ee
It is easy to see that $T(\tau,\rho)$ given by \eqref{sol1} fulfills this equation,
but now we can obtain also other solutions, in particular those for which 
$dT$ does not depend on time and the f-metric  is static. 
The most general solution of \eqref{sol11} of this type  
is obtained by separating the variables, 
\be                                  \label{static}
T=\sqrt{1+q^2}\,\tau+\int\frac{\rho \,d\rho}{\Sigma}\,\sqrt{q^2+\rho^2}\,,
\ee
where $q$ is an integration constant. If $q=0$ then the solution becomes 
especially simple, 
\be
T=\tau+\int\frac{d\rho}{\Sigma}-\rho\equiv V-\rho, 
\ee
and  choosing $V$ and $\rho$ as coordinates, the two metrics become 
\bea
ds_g^2&=&\alpha^2\{-\Sigma\, dV^2+2dV d\rho+\rho^2d\Omega^2\}, \nonumber \\
ds_f^2&=&u^2\alpha^2\{-dV^2+2dV d\rho+\rho^2d\Omega^2\}. 
\eea

\section{Energy}
One can compute the canonical energy for systems with non-trivial Stuckelberg fields
in the same way as this is done in the unitary gauge 
\cite{Volkov:2014qca,*Volkov:2014ida}.  
The computation will be presented separately \cite{MM} but its result is as follows. 
For a solution expressed in the static coordinates $\tau,\rho$ the energy 
on a hypersurface of constant $\tau$ is 
\be
E=\int {\cal E}\,d\rho
\ee
with the radial energy density  
\be
{\cal E}=u^2 P_2(u)\rho^2 \partial_\tau T. 
\ee
Applying this formula to the $T=t$ solution \eqref{sol1} gives 
the time-dependent value, 
\be
{\cal E}=u^2 P_2(u)\rho^2\sqrt{1-\rho^2}\cosh(\tau),
\ee
which indicates once again that this solution cannot describe the ground state.  
On the other hand, the energy will be time independent
if $\partial_\tau T$ is time independent,
but all such solutions are given by \eqref{static}, in which case  
\be
{\cal E}=u^2 P_2(u)\sqrt{1+q^2}\rho^2.
\ee
This corresponds to the constant volume energy density 
\be
\epsilon=u^2 P_2(u)\sqrt{1+q^2},
\ee 
and the 
total energy is $E=\epsilon V$ where $V$ is the (infinite) volume of the 3-space. 
We remember that $u$ is a solution of the algebraic equation $P_1(u)=0$, 
therefore, depending on choice of $u$ and also on values of the parameters $b_k$,
the energy can be positive, negative, or zero.  
 
The actual value of the background energy is probably not so important, 
but it is important to know if the energy is minimal or not. 
We conjecture that the static solutions \eqref{static} correspond to the energy minima and 
are therefore stable. Therefore, they are candidates for describing the de Sitter 
ground state in the theory.  To prove the conjecture will
require to resolve the constraints  and to compute the energy for deformations of the background 
\cite{Volkov:2014qca,*Volkov:2014ida}.  We presently have partial results 
supporting our conjecture, but the detailed analysis will be presented elsewhere \cite{MM}.

\section{Conclusions}

We have shown  that the de Sitter vacua in the dRGT theory are labeled by solutions of 
$(\partial_t{T})^2-(\partial_r T)^2=1$. The simplest solution $T=t$
is manifestly homogeneous and isotropic when written in the open chart, but it is unstable. 
Therefore, one should study other solutions. One could worry that other solutions
will not be FLRW because their reference metric is inhomogeneous. 
However, as we have seen, this is not 
necessarily the case, as  the reference metric can look inhomogeneous in some coordinates 
while sharing common translational isometries with the physical metric. 

The important issue is the number of common isometries of the two metrics. Since each 
of them describes a maximal symmetry space, each metric has ten isometries, some of which can be 
common, as for example the SO(3) rotational isometries.  The number of common isometries depends
on choice of $T$, for example for $T=t$ this number is  six, but the same 
can be true for other choices of $T$ as well. 

Requiring the timelike isometry to be common for both metrics reduces the 
set of solutions to a one-parameter family \eqref{static}. These solutions are physically 
distinguished since only for them the energy is time-independent. We conjecture that 
these solutions are stable and describe therefore the de Sitter ground state of the theory. 
The stability will follow if one shows that the energy increases for deformations 
of the de Sitter background, but such an analysis goes beyond the scope of the present paper 
and will be reported separately \cite{MM}.

\acknowledgments
We are grateful  to Gary Gibbons and Kei-ichi Maeda for constructive suggestions. 
The work of M.S.V. was partly supported by the Russian Government Program of Competitive Growth 
of the Kazan Federal University.

\appendix

\section{$T$-equation in general coordinates \label{ApA}}

\renewcommand{\theequation}{\Alph{section}.\arabic{equation}}
The most general spherically symmetric g-metric 
can be represented as 
\bea                             \label{dSg}
ds_g^2=\alpha^2\left\{-N^2dt^2+\frac{1}{\Delta^2}\,(dr+\beta\, dt)^2
+R^2\,d\Omega^2\right\},    \label{dsg} 
\eea 
where $N,\beta,\Delta,R$ are functions of two coordinates $t,r$. Choosing 
the f-metric to be  
\bea                             \label{dSgA}
ds_f^2=\alpha^2u^2\left\{-dT^2(t,r)+dR^2
+R^2\,d\Omega^2\right\}    \label{dSsg} 
\eea 
and analyzing the field equation $G_{\mu\nu}=T_{\mu\nu}$ component by component, 
one finds \cite{Volkov:2013roa} that they reduce 
to $G_{\mu\nu}+\Lambda g_{\mu\nu}=0$ with $\Lambda=P_0(u)$
provided that 
$P_1(u)=0$ and $\alpha^2=3/\Lambda$, and if  
$
(b_2+b_3 u)\,Y=0
$
hence either $b_2+b_3 u=0$ or $Y=0$. Here 
\bea                      \label{c0}
Y\equiv \left(\dot{T}-\beta{T}^\prime+N\Delta R^\prime \right)^2-
\left(\dot{R}-\beta{R}^\prime+N\Delta T^\prime \right)^2-
\left(
\Delta(\dot{T}R^\prime-\dot{R}T^\prime)+N
\right)^2\,,~~~
\eea
where the dot and the prime denote the derivatives with respect to $t$ and $r$, respectively. 

If the g-metric is de Sitter and 
$t,r$ coincide with the  $t,r$ coordinates  
of the ambient 5D Minkowski space used in \eqref{dS},
then one has 
\be                            \label{dSg1}
N=\frac{1}{\sqrt{1+t^2}}\,,~~~R=\sqrt{1+t^2-r^2}\,,~~~\Delta=NR\,,~~~\beta=-\frac{tr}{1+t^2}\,.
\ee
Inserting this to the condition $Y=0$ gives equation \eqref{T} for $T(t,r)$ in the main text. 
The other possibility 
is to set $b_2+b_3 u=0$ \cite{Chamseddine:2011bu},%
\cite{Kobayashi:2012fz,*Kodama:2013rea}, which restricts the  
values of the parameters $b_k$ but the function $T(t,r)$ then remains arbitrary, 
which presumably indicates some hidden gauge invariance.  


%

\end{document}